\documentclass[twocolumn,showpacs,showkeys]{revtex4}
\usepackage{bm}% bold math
\usepackage{graphics}
\usepackage{epsfig}
\usepackage{color}
\usepackage{times}
\usepackage[ps2pdf,colorlinks=true, pagebackref=false, bookmarks=true,bookmarksopen=true,bookmarksnumbered=true]{hyperref}

\newcommand{\Cu}{\ensuremath{\mathrm{Cu}}}
\newcommand{\Ni}{\ensuremath{\mathrm{Ni}}}
\newcommand{\Nb}{\ensuremath{\mathrm{Nb}}}
\newcommand{\Al}{\ensuremath{\mathrm{Al}}}

\renewcommand{\O}{\ensuremath{\mathrm{O}}}
\renewcommand{\S}{\ensuremath{\mathrm{S}}}
\newcommand{\F}{\ensuremath{\mathrm{F}}}
\newcommand{\Ar}{\ensuremath{\mathrm{Ar}}}
\newcommand{\Y}{\ensuremath{\mathrm{Y}}}
\newcommand{\Ba}{\ensuremath{\mathrm{Cu}}}
\newcommand{\C}{\ensuremath{\mathrm{C}}}

\begin{document}

\title{%
Low-$T_c$ Josephson junctions with tailored barrier
}

\author{M. Weides}
\email{m.weides@fz-juelich.de}
\author{C. Schindler}
\author{H. Kohlstedt}
\affiliation{Institute for Solid State Research (IFF) and
  Center of Nanoelectronic Systems for Information Technology (CNI), Research Centre Juelich, D-52425 Juelich, Germany}

\date{\today}% It is always \today, today,
             % but any date may be explicitly specified

\begin{abstract}
$\Nb/\Al_2\O_3/\Ni_{0.6}\Cu_{0.4}/\Nb$ based superconductor-insulator-ferromagnet-superconductor (SIFS) Josephson tunnel junctions with a thickness step in the metallic ferromagnetic $\Ni_{0.6}\Cu_{0.4}$ interlayer were fabricated. The step was defined by optical lithography and controlled etching. The step height is on the scale of a few angstroms. Experimentally determined junction parameters by current-voltage characteristics and Fraunhofer pattern indicate a uniform F-layer thickness and the same interface transparencies for etched and non-etched F-layers. This technique could be used to tailor low-$T_c$ Josephson junctions having controlled critical current densities at defined parts of the junction area, as needed for tunable resonators, magnetic-field driven electronics or phase modulated devices.
\end{abstract}

%  74.50.+r,%Proximity effects, weak links, tunneling phenomena,
           %and Josephson effect
%  85.25.Cp %Josephson devices
% 75.70.cn Magnetic properties of interfaces (multilayers, superlattices, heterostructures)
% 74.20.Rp %Pairing symmetries (other than s-wave)
%  74.25.Ha Magnetic properties
% 74.78.Db %Low-Tc films
% 75.45.+j,%Macroscopic quantum phenomena in magnetic systems
%  74.78.Fk %Multilayers, superlattices, heterostructures
%  74.81.-g %Inhomogeneous superconductors and superconducting systems
% 74.81.Fa Josephson junction arrays and wire networks
%  85.25.Dq Superconducting quantum interference devices (SQUIDs)
%  74.45.+c %Proximity effects; Andreev effect; SN and SNS junctions



\maketitle

\section{Introduction}

\begin{figure}[tb]
\begin{center}
\includegraphics[width=8.6cm]{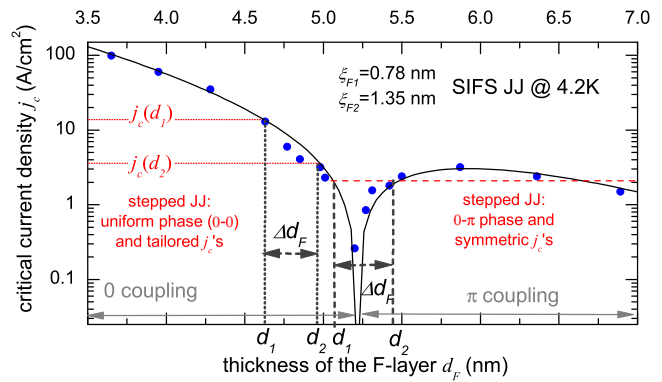}
 \caption{(color online) $j_c(d_F)$ dependences and fitting curve for SIFS-JJs. F-layer thicknesses chosen as $d_1$ and $d_2$ in stepped JJs yields $0$--$0$ coupled junction with asymmetry in $j_c$ (dotted lines) or symmetric $0$--$\pi$ junction (dashed lines). Data from Ref. \cite{WeidesHighQualityJJ}. \label{IcdFMod}}
\end{center}
\end{figure}

The work horse in superconducting electronics is the \emph{Josephson junction} (JJ). A Josephson junction consists of two weakly coupled superconducting metal bars via a constriction, e.g. made up by a normal (N) metal or a tunnel barrier (I). Various types of Josephson junctions are routinely applied in ultra-high sensitive SQUID (Superconducting Quantum Interference Devices) magnetometers, radio astronomy receivers or the voltage standard \cite{BuckelKleiner2004Superconductivity}. Especially $\Nb/\Al-\Al_2\O_3/\Nb$ low-$T_c$ tunnel junctions attract considerable interest in many respects. The $\Al$ over layer technique \cite{Gurvitch82NbAlONb} allows the fabrication of high density $\Nb$-based Josephson circuits with compromising small parameters spreads. Nonetheless, the Josephson junction itself is a research subject which enriches our understanding of superconductivity, transport phenomena across interfaces and tunnel barriers. Here it is  worthwhile to note, that with the advent of high-quality magnetic tunnel junctions approximately 10 years ago, new so far unexplored devices are now under development which make use of both fabrication techniques and which consists of advanced layer sequences compromising superconducting (S) and magnetic materials (F).\\At superconductive/magnetic metal (S/F) interfaces the superconducting order parameter $\Psi$ is \emph{spatially decaying and oscillating} inside the magnet (coherence length $\xi_{F1}$, oscillation length $\xi_{F2}$), whereas for a S/N system $\Psi$ is simply \emph{decaying} inside the metal \cite{buzdin05RMP}. By combining the low-$T_c$ $\Nb/\Al$ technology with magnetic tunnel junctions new functionalities are predicted. In this framework so called $0$--$\pi$ Josephson junctions were recent focus of research activities \cite{RoccaAprili05classicalspins,FrolovRyazanovSemifluxon,WeidesFractVortex}.\\
The supercurrent through an SNS junction is given by $I=I_c\sin(\phi)$, where $\phi=\Psi_1-\Psi_2$ is the phase difference of the superconducting electrode wave functions and $I_c>0$ the maximum supercurrent through the junction \cite{Josephson}. In absence of current ($I=0$) through the JJ the Josephson phase $\phi=0$ corresponds to the energy minimum. These junctions are so-called $0$ JJs. In an SFS stack with ferromagnetic layer thickness $d_F\propto \xi_{F2}/2$, the amplitude of order parameter $\Psi$ vanishes at the center of the F-layer and the order parameter has the opposite sign at the adjacent superconducting electrode. This state is described by a phase shift of $\pi$ and these junctions are so-called $\pi$ JJs. SFS-type $\pi$ JJs have a negative critical current, hence the Josephson relation can be rewritten: $I=-I_c\sin(\phi)=|I_c|\sin(\phi+\pi)$ \cite{Bulaevskii1977,Buzdin1991}. Recently, these types of Josephson junctions have been realized using SFS \cite{Ryazanov2000,Blum:2002:IcOscillations} and SIFS \cite{Kontos02Negativecoupling,WeidesHighQualityJJ} stacks. The kind of coupling can be determined by the $j_c(d_F)$ dependence, see Fig. \ref{IcdFMod}.

\iffalse The difference in spin ordering leads to a very rich and interesting physics.\fi
\iffalse Superconductive (S), non-magnetic (N) or ferromagnetic (F) metals are well-known physical properties of solid states. While the interface physics of superconductivity/non-magnetic metal (S/N) has been widely studied for a long time\cite{deGennes},\fi
%\subsubsection*{Josephson junction with tailored $j_c$}

For a variety of Josephson junctions a non-uniform critical current density $j_c$ is desirable, as for example for tunable superconducting resonators, toy systems for magnetic flux pinning or magnetic-field driven electronic switches similar to SQUIDs.
The first considerations \cite{Russo1978NonuniformJc} of non-uniform $j_c$'s were caused by technological drawbacks leading to variations of barrier thicknesses by fabrication \cite{Schwidal1969} or of illumination in case of light-sensitive junctions \cite{Barone1977PhysStat}. Later the properties of JJs with periodic spatially modulations were intensively studied regarding the pinning of fluxons \cite{McLaughlinScottPRB1978,Vystavkin1988,MaloUstinovJAP1990}, the spectrum of electromagnetic waves \cite{FistulPRB1999,Lazarides2005} or their magnetic field dependences \cite{LazaridesPeriodicDefects2003}. Experimentally the spatial modulation of $j_c$ was realized lithographically
by inserting of artificial defects such as insulation stripes across the barrier (ex-situ layer process, $j_c=0$) \cite{GolubovUstinovPLA1988,ItzlerTinkhamPRB95}, microshorts ($j_c$ increased) or microresistors ($j_c$ decreased). The properties of JJs depend on geometrical (width, length, thickness) and the physical
(dielectric constant of insulator $\epsilon$, resistance $\rho$, magnetic thickness $\Lambda$ and $j_c$) parameters. When tailoring $j_c$ all other parameters should be unchanged to facilitate calculations and avoid further inhomogeneities in the system. The conventional methods for changing $j_c$ intrinsically modify either $\epsilon$ or $\rho$, too. Our fabrication technology permits the controlled change of only the interlayer thicknesses $d_1$ and $d_2=d_1+\Delta d_F$, i.e. the local $j_c$.\par

%\subsubsection*{$0$--$\pi$ Josephson junctions}\label{0piJJ}
The case of non-uniform coupling phase within a \emph{single}
Josephson junction, i.e. one half is a $0$ JJ ($d_F=d_1$) and the other half is a $\pi$ JJ ($d_F=d_2$) (see dashed lines in Fig. \ref{IcdFMod}) is of particular interest.
In such a $0$--$\pi$ junction a spontaneously formed vortex of supercurrent circulating around the $0$--$\pi$ phase boundary with flux $|\Phi|\leq\pm\Phi_0/2$ inside the JJs may appear \cite{Bulaevskii1978}. The sign of flux depends on the direction of the circulation and its amplitude equal to $|\Phi_0/2|$, i.e. a semifluxon, if the junction length $L$ is much larger than the Josephson penetration depth $\lambda_J$ \cite{Xu:SF-shape,Goldobin02SF}. The ground state depends on the symmetry ratios of critical currents $|j_c(d_1)|/|j_c(d_2)|$ and the effective junction lengths $\ell_{1}/\ell_{2}$ of $0$ and $\pi$ parts ($\ell=L/\lambda_J$). The $0$--$\pi$ junctions
have been actively studied both in theory and experiment during the
past few years \cite[and references
herein]{Tsuei:Review,Kirtley:IcH-PiLJJ,Hilgenkamp:zigzag:SF,Goldobin03GS_biasReArrange,ZenchukGoldobin04GS,Goldobin:2005:QuTu2Semifluxons,WeidesFractVortex}.
\iffalse
The ideal $0$--$\pi$ JJ would have equal
$|j_c(0)|=|j_c(\pi)|$ and $0$--$\pi$ phase boundary in its center to have the
symmetric situation. Furthermore the junctions should be underdamped (SIFS
structure) since low dissipation is necessary for the study of dynamics.\par
In these systems the phase shift of $\pi$ is given by the anisotropic order parameter of the $d$-wave superconductor and not at the barrier.\fi
$0$--$\pi$ junctions were studied at so-called tricrystal grain boundaries in $d$-wave superconductors \cite{Tsuei:Review}, later in $\Y\Ba_2\Cu_3\O_7$--$\Nb$ ramp zigzag junctions \cite{Hilgenkamp:zigzag:SF} and $\Nb$ based JJ using current injectors \cite{Goldobin04PRLDynamicsSF}. The advantage of SFS/SIFS technology over these systems is that by a proper chosen F-layer thickness $d_F$ the phase can be set to $0$ ($d_1$) or $\pi$ ($d_2$)
and the amplitude of the critical current densities
$j_c(0)$ and $j_c(\pi)$ can be controlled to some degree. It can be prepared in a multilayer geometry (thus allowing topological freedom of design), can be easily combined with the
well-developed $\Nb/\Al$--$\Al_2\O_3/\Nb$ technology and has good scalability.
\iffalse The fabrication of $0$ and $\pi$ JJs on the same chip based on
SFS/SIFS technology, e.g. in a dc-SQUID with $0$ and $\pi$ junctions was
reported \cite{GuichardApril03DCSquid}.\fi For $0$--$\pi$ junctions one
needs $0$ and $\pi$ coupling in \emph{one} junction, setting high demands on
the fabrication process, as the change of coupling demands exact control in F-layer thickness. $0$--$\pi$ JJs have been realized in SFS-like systems
 \cite{RoccaAprili05classicalspins,FrolovRyazanovSemifluxon}. However, both systems have the disadvantages that the $0$--$\pi$ phase boundary was prepared in an uncontrolled manner and does \emph{not} give information about $j_c$ in $0$ and $\pi$ coupled parts. Hence, the ratios of $|j_c(0)|/|j_c(\pi)|$ and $\ell_0/\ell_\pi$ cannot be calculated for these samples and their ground state is unknown.\par
In a recent publication \cite{WeidesFractVortex} the
authors presented the first \emph{controllable} stepped $0$--$\pi$ JJ of
SIFS type that are fabricated using high quality $\Nb/\Al_2\O_3/\-\Ni_{0.6}\-\Cu_{0.4}/\Nb$ heterostructures. These stepped junctions came along with reference junctions to calculate the ground state of $0$--$\pi$ JJs. \iffalse Generally, SIFS junctions pose the advantage over SFS JJs that i) the essential electric parameters, such as $j_c$, can be varied over a wide range by changing the tunnel barrier I thickness and ii) high $I_cR$ products are obtained.\fi
The requirements for SIFS $0$--$\pi$ junctions are challenging. Here we present our technology background for stepped junctions with the focus on small parameter spreads. Our approach represents a considerable step forward to fulfil the extreme requirements on the implementation of conventional or quantum computing devices based on Josephson junctions.

The current and/or coupling phase profile can be modified by a tailored \emph{stepped} barrier in SXS/SIXS-type (X=N, F) Josephson junctions. The patterning concept for stepped JJs presented in this article can be either used for JJ with tailored $j_c$'s and uniform phase $0$--$0$ ($\pi$--$\pi$) JJs (dots in Fig. \ref{IcdFMod}) or with symmetric $j_c$'s and non-uniform phase, i.e. $0$--$\pi$ JJs (dashes in Fig. \ref{IcdFMod}).

\iffalse These new class of low-$T_c$ JJs can be used for numerous kind of devices.\fi

\section{Experiment}

%\subsection{SIFS stack}
\begin{table}[tb] \begin{center}
\caption{Deposition (DC-sputtering) and etching parameters for SIFS stacks. The rates were determined by profiler measurements.\label{sputter}}
\begin{tabular}{ccccc}
  \hline
  \hline
  metal & $\Ar$ pressure  & power density  & rate & $\quad$ parameters$\quad$\\
  &[$10^{-3}\:\rm{mbar}$]&[$\rm{W/cm^2}$] & [$\rm{nm/s}$]& \\
  \hline
  $\Nb$ & $7.0$ & $5$ & $2.0$& static\\
  $\Al$ & $7.0$ & $1.9$ & $0.05$& rotation\\
  $\Ni\Cu$ & $4.2$ & $0.6$ & $\leq0.34$& target shifted\\
  $\Cu$ & $4.2$ & $1.9$ &$0.1$& rotation\\
   \hline
  $\S\F_6$ on \Nb & $15$ & $0.6$ &$\sim1$ & RF-source\\
  $\S\F_6$ on \Ni\Cu & $15$ & $0.6$ & $<$0.001 & RF-source\\
  $\Ar$ on \Ni\Cu & $5$ & $0.6$ &$\sim0.01$ & RF-source\\
   \hline
  \hline
\end{tabular}
\end{center}
\end{table}

A Leybold Univex 450b magnetron sputter system with 8 targets in the main chamber, a load-lock including an etching stage, as well as a separate oxidation chamber was used for the junction preparation. Together with a transfer chamber, a robot handler and a Siemens Simatic control unit, the cluster tool is able to deposit automatically (pre-programmed) tunnel junction layer sequences for superconducting and spintronics applications.\\
The deposition and patterning of the stepped SIFS junctions was performed by a four level
photolithographic mask procedure. Here we present an improved version of an earlier fabrication sequence for planar SIS junctions \cite{Gurvitch82NbAlONb}. The magnetron sputtering system was capable of handling $4$--inch wafers and had a background pressure of $5\cdot10^{-7}\;\rm{mbar}$ was used. $\Nb$ and $\Ni\Cu$ were statically deposited, while $\Al$ and $\Cu$ were deposited during sample rotation and at much lower deposition rates to obtain
very homogeneous and uniform films, see table \ref{sputter}. Although the actual stack sequence for the junctions is SINFS, the N-layer ($\Cu$) was introduced to provide the growth of a uniform and homogenous F-layer thickness \cite{WeidesFabricationJJPhysicaC}. It is not relevant for the electric and magnetic properties discussed here and will be neglected.\\
The F-layer was deposited with a gradient of thickness along $y$-axis on the S/I stack \cite{WeidesFabricationJJPhysicaC}. The increase of
thickness over the junctions width ($\leq100\;\rm{\mu m}$) is estimated as less than $0.02\;\rm{nm}$, i.e. the F-layer can be treated as planar for an individual junction. After the deposition of $40\;\rm{nm}$ $\Nb$
as cap layer and subsequent lift-off the complete SIFS stack with wedge-shaped
F-layer thickness, but without steps in F-layer, was obtained.

%\subsection{Patterning of step}
\begin{figure}[tb]
\begin{center}
\includegraphics[width=8.6cm]{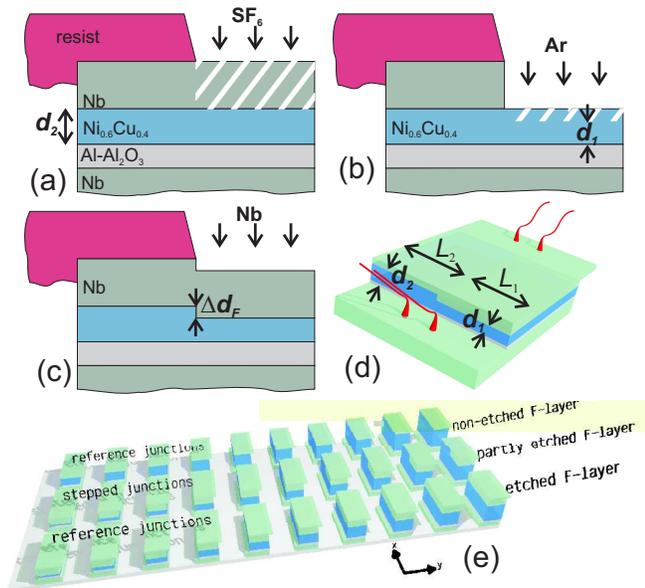}
\caption{\label{SFpatterning} (color online) The complete SIFS stack was protected in part by photoresist.  The
$\Cu$-layer was necessary for uniform current transport: (a) reactive etching of
$\Nb$ with $\S\F_6$ down to $\Ni\Cu$ layer, (b) ion-etching of $\Ni\Cu$ to set $0$
coupling, (c) in situ deposition of cap $\Nb$ layer. Schematic layouts of
stepped JJ based on SFS/SIFS technology (d) and of stepped JJ along with planar reference junctions (e). The F-layer (blue) thickness increases from left to right.}
\end{center}
\end{figure}

The patterning of the desired step-like variation in $d_F$ was done after the complete deposition of the SIFS stack. The parts of the JJ that were supposed to have a larger
thickness $d_2$ were protected by photoresist, see Fig.~\ref{SFpatterning}.
Ion-etching alone of both $\Nb$ and $\Ni\Cu$ to define the step in F-layer did not provide a good control over the final F-layer thickness, as this unselective and long-timed etching has
the disadvantages of non-stable etching rates and an non-uniform etching front.
Therefore it was not possible to achieve in such way a defined step.\par
The use of selective etching in $\Nb/\Al$--$\Al_2\O_3/\Nb$ stack fabrication processes, such as $\C\F_4$ and $\S\F_6$ reactive ion etching (RIE) or other techniques are reported in Ref.~ \cite{SF6EtchingLichtenberger1993}. In particular, it was shown that $\S\F_6$ provides an
excellent RIE chemistry for low-voltage anisotropic etching of $\Nb$ with high selectivity towards
other materials. The inert $\S\F_6$ dissociated in a
RF-plasma and the fluor diffused to the surface of the substrate, where it reacted with niobium
$5\F+\Nb\longrightarrow\Nb\F_5$. The volatile $\Nb\F_5$ was pumped out of the etching-chamber.
When $\S\F_6$ was used as process gas all non-metallic etching products such as fluorides and sulfides from the top-layer of the $\Ni\Cu$-layer had to be removed by subsequent argon etching.\par

The patterning process of the step is depicted in Fig.~\ref{SFpatterning} (a)--(c). The key points were i) \emph{selective reactive etching} of $\Nb$, ii) \emph{argon etching} of $\Ni\Cu$ to define $d_1=d_2-\Delta d_F$ and iii) subsequent \emph{in situ} deposition of $\Nb$.

%\subsubsection*{step}
The $\Nb$ cap layer was removed by reactive dry etching using $\S\F_6$ with a high selectivity to the photoresist (AZ5214E). A few tenth of nanometer $\Delta d_F$ of $\Ni\Cu$ were $\Ar$ ion etched
at a very low power and rate to avoid any damaging of the $\Ni\Cu$ film under
the surface and to keep a good control over the step height. When the F-layer
thickness was reduced down to the thickness $d_1$ the etching was stopped and
$40\;\rm{nm}$ of $\Nb$ were deposited. The complete etching and subsequent $\Nb$ deposition was done in situ at a background pressure below $2\cdot 10^{-6}\;\rm{mbar}$. The chip contained stacks with the new F-layer thicknesses $d_1$ (uniformly etched), $d_2$ (non-etched)
and with step in the F-layer thickness from $d_1$ to $d_2$.

%\subsection{Final patterning}
After the preparation of steps in F-layer the actual junction areas were defined by
aligning the photo mask on the visible step-terraces (ramp of $\leq20 \;\rm{nm}$ height and
$\sim1 \;\rm{\mu m}$ width), followed by $\Ar$ ion-beam etching of the upper $\Nb$, $\Ni\Cu$ and $\Al$ layers. The length $L_1$ and $L_{2}$ of a stepped junction are within lithographic alignment accuracy of $\sim1\;\rm{\mu m}$. The etching was controlled by a secondary
ion mass spectrometry (SIMS) and stopped after the complete etching of the $\Al_2\O_3$ tunnel barrier. Afterwards the mesas were insulated by SNEAP (Selective Niobium Etching and Anodization Process) \cite{Gurvitch82NbAlONb}. In the last photolithographic step the wiring layer was defined. After a short argon etching to reduce the contact resistance a $300 \: \rm{nm}$ thick $\Nb$ wiring was deposited. Fig.~\ref{SFpatterning} (d) sketches a stepped SIFS junction and (e) the completely structured chip with sets of stepped and planar reference junctions and wedge shaped F-layer along $y$-axis.
Fig. \ref{IcdFMod} depicts the $I_c(d_F)$ dependence of planar SIFS JJs with $\Ni\Cu$ as ferromagnetic interlayer. Note that a simple decay of $I_c$ can generally be achieved by increasing the interlayer thickness, independent of its magnetic properties, i.e. in SNS or SINS junctions. The wedge shaped interlayer (Fig. \ref{SFpatterning} (e)) facilitates the quick fabrication of samples with various F-layer thickness and, at the same time, a low junction to junction deviation.

\section{Results and Discussion}

\begin{figure}[tb]
\begin{center}
\includegraphics[width=8.6cm]{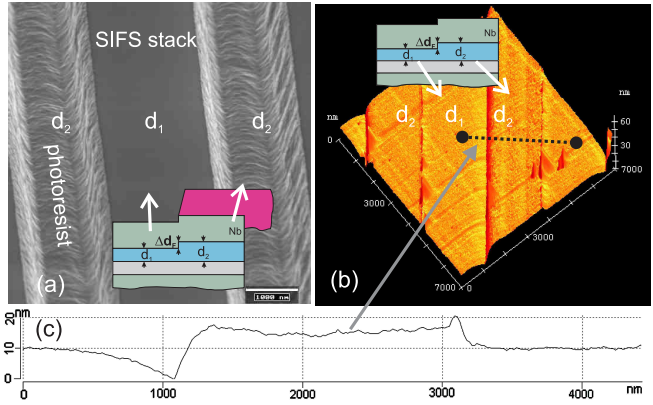}
\caption{(color online) Topography of test sample with multiple steps and $2\;\rm{\mu m}$ wide gaps after
etching and $\Nb$ deposition. (a) SEM image before removal of photoresist (protecting $\pi$ coupled parts); (b) AFM image ($7\times7\;\rm{\mu m^2}$) after removal of photoresist; (c) Profile measured by AFM (dotted line).
\label{SEMAFM}}
\end{center}
\end{figure}

On test samples multiple steps with $2\;\rm{\mu
m}$ gaps were structured for analysis with scanning electron (SEM) and atomic force (AFM) microscopy after etching and $\Nb$ deposition, see Fig.~\ref{SEMAFM}.
For SEM the photoresist was left on the sample and for AFM it was removed. While etching with $\S\F_6$ the $\Ni\Cu$-layer served as an etching
barrier. Thus, it facilitated the over-etching of the $\Nb$ to ensure its
complete removal despite the shadow effects from resist walls ($\sim1\;\rm{\mu m}$ height). The short-timed argon etched $\Ni\Cu$-layer was slightly
non-uniform near the resist walls due to the
anisotropic etching front. Since in real stepped JJs the half with $d_F=d_1$
has dimensions about $10\;\rm{\mu m}$ or larger, the non-uniformity of the
$\Ni\Cu$ layer near the step, created by shadow-effects of the
resist, was averaged out in transport measurements.
The presence of resist caused a decrease of the $\Nb$ deposition rates by $\sim10\%$, especially near the asymmetric resist walls (seen in SEM). Thus, the non-uniform deposition of $\Nb$ after the reactive and argon etching led to a non-uniform cap layer, seen in the difference of stack heights (much larger than $\Delta d_F$) in the AFM image of the SIFS stack, Fig. \ref{SEMAFM}(b).

On planar SIFS-JJs (reference junctions) the actual step height $\Delta d_F$ and the coupling is estimated by comparing $j_c(d_1)$ with the known $j_c(d_F)$ dependence, see Fig. \ref{IcdFMod}. $d_2$ is determined from a reference sample with wedge-shaped F-layer. \iffalse The
polycrystalline structure of the $\Ni\Cu$ film and the inherent interlayer roughness of
room-temperature sputtered films prevent the exact estimation of the layer thickness. \fi Our experimental results suggested that a continuously variable-thickness model was
more suitable for the junctions than a monolayers-thickness model (radius of neutral $\Ni$, $\Cu$ is $\sim0.15\;\rm{nm}$) \cite{WeidesPhD}. \iffalse The mean F-layer thickness $d_F$ of the Josephson junction was calculated by the sputter rates, being a function of the junction location along $x$-axis.\fi\\
By the IVC and $I_c(H)$ of the reference JJs one can estimate parameters for the stepped junction, such as the ratio of asymmetry $|j_c(d_1)|/|j_c(d_2)|$ and the quality of the etched and non-etched parts. The uniformity of the supercurrent transport in a Josephson junction can be judged qualitatively from the magnetic field dependence of the critical current $I_c(H)$. The magnetic field $H$ was applied in-plane and along one junction axis. The magnetic diffraction pattern depends in a complex way on the effective junction length $\ell$ and on the current distribution over the junction area \cite{BaronePaterno}. The ideal pattern of a short ($\ell\leq1$) JJ is symmetric with respect to both polarities of the critical
current and the magnetic field with completely vanishing  $I_c$ at the minima. If the pattern is not symmetric, irregular or has a current offset, the current transport over the non-superconducting interlayers is non-uniform. If the trapping of magnetic flux can be excluded this effect is attributed to non-uniformity of the tunnel barrier, the ferromagnetic layer or the interface transparencies over the junction area.\par

\subsection{Reference junctions}
\begin{figure}[tb]
\begin{center}
  \includegraphics[width=8.6cm]{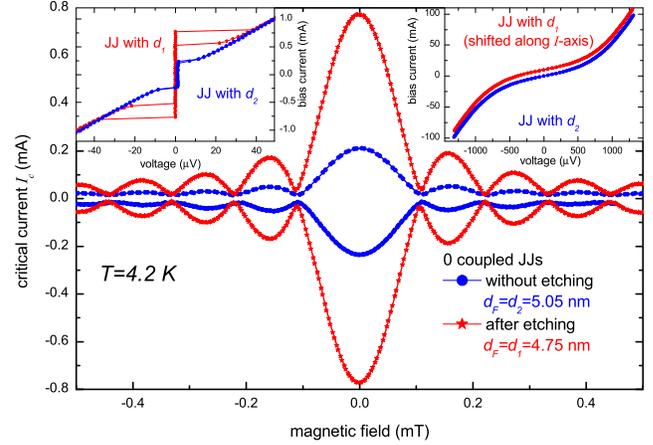}
  \caption{(color online) $I_c(H)$ of etched (star) and non-etched (dots) JJs. The insets show IVCs for small and large bias current ranges in zero magnetic field. Both JJs are in the short JJ limit. Measurements were done at $4.2\;\rm{K}$.}
  \label{Fig:IcHIVtogether}
\end{center}
\end{figure}

Fig.~\ref{Fig:IcHIVtogether} shows IVC and $I_c(H)$ dependences for a non-etched
junction (dot) with the F-layer thickness $d_2=5.05\;\rm{nm}$ and a uniformly etched junction
(star) with thickness $d_1=d_2-\Delta d_F=4.75\;\rm{nm}$. Both junctions were $0$ coupled. The step height $\Delta d_F=0.3\;\rm{nm}$ is calculated by comparing $I_c(d_1)$ with $j_c(d_F)$ from Fig. \ref{IcdFMod}. The polycrystalline structure of the room-temperature sputtered layers and the very low etching rate of $\Ni\Cu$ led to a good control over $\Delta d_F.$ However, one has to keep in mind that the local variation of F-layer thickness might exceed this value, and the values for $d_1$, $d_2$ and $\Delta d_F$ are just the mean thicknesses seen by the transport current. The insets of Fig.~\ref{Fig:IcHIVtogether} show the IVCs for small and large ranges of bias current. Besides the difference in $I_c$, the $I_c(H)$ dependence and the
IVCs (right inset) showed no evidence for an inhomogeneous current transport
for both samples. The larger $I_c$, but same resistance $R$ and capacitance $C$ led to
a slightly hysteretic IVC of the etched sample, as the width of hysteresis is determined by McCumber parameter $\beta_c \propto I_c^2 R C$. The resistance $R$ is nearly independent from $d_F$ as the voltage drop over the tunnel barrier is much larger than the serial resistance of a few nanometer thick metal \cite{WeidesHighQualityJJ}. However, an etching-induced change of transparency at the F/S interface might modify $R$. No change of $R$ is visible in the IVCs of both JJs in Fig.~\ref{Fig:IcHIVtogether}, apart from the change in $I_c$. A change of capacitance $C$ would require a change of $R$, as both are determined by the dielectric tunnel barrier. \par

The scattering of the critical current $I_c$ and resistance $R$ on the etched junctions was, just
like for the non etched SIFS junctions, of the order of $2\%$. The trapping of magnetic flux might cause a larger variation in critical current density than the interface asymmetry stemming from the etching process. For all thicknesses of ferromagnetic layer the same homogeneity of the etched junctions were observed.\par

A set of junctions denoted by the dashes in Fig. \ref{IcdFMod} was measured by the authors, too. The $I_c(H)$ dependences of this $0$, $\pi$ and $0$--$\pi$ JJs are depicted in Fig.~3 of Ref. \cite{WeidesFractVortex}. These junctions show the same quality of parameters for the etched and non-etched samples as the samples depicted in  Fig. \ref{Fig:IcHIVtogether}.

\subsection{Stepped junctions}

\subsubsection{Calculated $I_c(h)$ of stepped JJ}
\begin{figure}[tb]
\begin{center}
  \includegraphics[width=8.6cm]{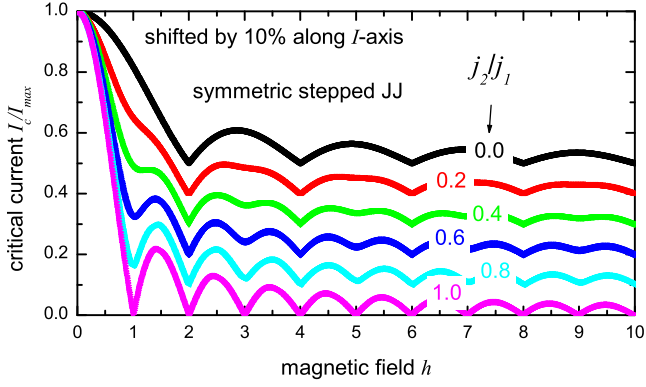}
  \caption{(color online) Calculated $I_c(h)$ dependence for various ratios of $j_2/j_1$ and centered step in $j_c$ profile.}
  \label{Fig:simuSteJJIcH}
\end{center}
\end{figure}

The magnetic diffraction pattern $I_c(H)$ of a JJ depends on its $j_c$ profile, see Ref.~\cite{BaronePaterno,Note0Pi}. The analytic solution for a short ($\ell<1$) stepped junction with different critical current density $j_1$ and $j_2$ in both halves is given by
\[\frac{I_c(h)}{A}=\frac{j_1\cos{(\phi_0-h)}-j_{2}\cos{(\phi_0+h)}+(j_2-j_1)\cos{\phi_0}}{2h},\]
where $\phi_0$ is an arbitrary initial phase, $h=2\pi\Lambda\mu_0LH/\Phi_0$ the normalized magnetic flux through the junction cross section, $\Lambda$ the magnetic thickness of junction and $A$ the junction area. The phase-field relation for maximum $I_c$ is reached for
\[\phi_0=\arctan{[\frac{j_{2}h\sin{h}-j_1h\sin{h}}{2hj_{2}\sin^2\left(\frac{h}{2}\right)+2hj_{1}\sin^2\left(\frac{h}{2}\right)}
].}\]\\
The calculated $I_c(h)$ for various ratios of $j_2/j_1$ is depicted in Fig.~\ref{Fig:simuSteJJIcH}. Characteristic features are the centered maximum peak and the appearance of periodic minima of the supercurrent for $h=n$ for integer $n$. The height of the odd-order minima ($n=1,3,5...$) depends on the asymmetry ratio $j_c(d_2)/j_c(d_1)$ and increases for decreasing $j_c(d_2)$. $I_c(h)$ is completely vanishing for magnetic flux equal to multiples of $2\Phi_0$. The maximum critical current at $I_c(0)$ decreases linearly down to $I_c/I_{\rm{max}}=0.5$ for $j_c(d_2)=0$. The corresponding $I_c(h)$ pattern becomes that of a junction with half the width and uniform $j_c$.

\subsubsection{Measured $I_c(H)$ of stepped JJ}

\begin{figure}[tb]
\begin{center}
  \includegraphics[width=8.6cm]{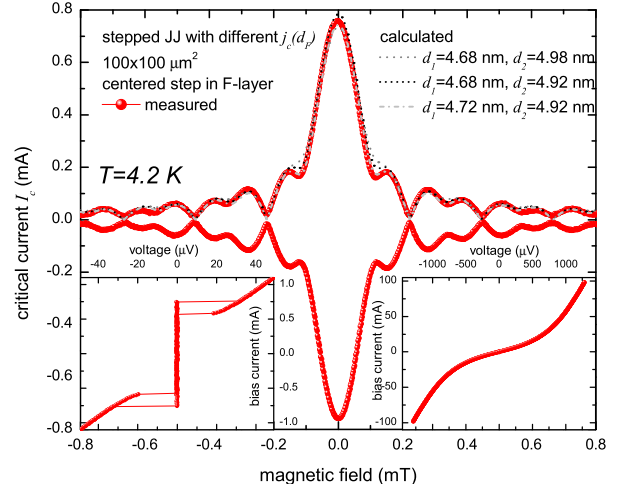}
  \caption{(color online) $I_c(H)$ of a stepped $0$--$0$ JJs (square shaped with $100\;\rm{\mu m}$ junction length) with $d_1=4.68\;\rm{nm}$ and $d_2=4.98\;\rm{nm}$ (determined from reference JJs) plus calculated $I_c(h)$. The junction is in short JJ limit.}
  \label{Fig:IcHAsym}
\end{center}
\end{figure}

In Fig.~\ref{Fig:IcHAsym} the measured magnetic diffraction pattern $I_c(H)$ of a stepped JJ along with calculated $I_c(h)$ curves are depicted. Both junctions halves are $0$ coupled. The magnetic field axis $h$ was scaled to fit the first measured minima of $I_c(H)$. $d_F$ is determined by comparing $j_c$ of reference junction with the known $j_c(d_F)$ dependence in Fig. \ref{IcdFMod}, yielding $d_1=4.68\;\rm{nm}$ and $d_2=4.98\;\rm{nm}$. Due to the rather steep slope of the $j_c(d_F)$ curve near the $0$ to $\pi$ crossover at $d_F=5.21\;\rm{nm}$, $j_c$ is very sensitive to $d_1$ and $d_2$. A variation of $d_1$ and $d_2$ by $0.05\;\rm{nm}$ changes $j_c$ up to $30\%$. The sputter rate may vary slightly over the $\sim500\;\rm{\mu m}$ distance to the reference junctions, preventing the exact estimation of $d_1$ and $d_2$.  \iffalse The parameters used for the calculation are the $j_c(d_F)$ dependence from Fig.~\ref{IcdFMod}, the thickness $d_2$ and the step height $\Delta d_F$.\fi We calculated $I_c(h)$ (dashes) using $d_1=d_2-\Delta d_F=4.68\;\rm{nm}$ and $d_2=4.98\;\rm{nm}$ determined from reference junctions. Then $d_2$ was decreased by $0.04\;\rm{nm}$ (dots) and finally $\Delta d_F$ increased by the same thickness, too (dashes-dots). The final calculation has the best agreement with data, although the total interface roughness (rms) of the multilayers should exceed $0.04\;\rm{nm}$ by far. The measurement and simulation in Fig.~\ref{Fig:IcHAsym} show the good estimation and control of F-layer thicknesses.

\section{Conclusions}
Josephson junctions with a step in the ferromagnetic layer were fabricated. Using a wedge-shaped F-layer in a SIFS stack on a $4$--inch wafer along with stepped and reference junctions it was possible to trace out regimes of different couplings ($0$, $\pi$), depending on the initial F-layer thickness $d_2$ and step $\Delta d_F$. The etched and non etched SIFS junctions differ only by F-layer thickness. No inhomogeneities can be seen in the current transport characteristics of the etched junctions.\\
The patterning of stepped JJs allows free lateral placement of well-defined $j_c$'s and/or local coupling regimes within a single junction. If decreasing temperature the slope $\partial j_c/\partial T$ depends on the interlayer thickness (observed in SIFS JJs \cite{WeidesHighQualityJJ}), and the ratio $j_1/j_2$ could be varied thermally. Stepped junctions can be realized in $\Nb$ based JJs with any interlayer material (N,F,I) which is chemically stable towards the reactive etching gas. The patterning process could be adjusted to all thin film multilayer structure providing that the reactive etching rates of the layer materials differ. Replacing the optical lithography with electron beam lithography may enhance the lateral accuracy of the step down to the dimension of e-beam and decrease the non-uniformity near the resist wall due to the thinner resist height.\\ JJs with varying $j_c$ and planar phase could be used for devices with special shaped $I_c(H)$ pattern \cite{BaronePaterno}, toy systems for flux pinning or tunable superconducting resonators. $0$--$\pi$ JJs based on low-$T_c$ superconductors with stepped F-layer offer a great flexibility for the integration of this devices, as it offers advantages over the existing $0$--$\pi$ junctions based on d-wave
superconductors \cite{Hilgenkamp:zigzag:SF,KirtleyImaging96PRL} or current injectors \cite{Goldobin04PRLDynamicsSF} such as the low dissipation of plasma oscillations, no restrictions in topology, no additional bias electrodes and easy integration into the mature $\Nb/\Al-\Al_2\O_3/\Nb$ technology.\\
The $0$--$\pi$ SIFS JJs with stepped F-layer allow to study the physics of fractional vortices with a good control of the ratio of symmetry between $0$ and $\pi$ parts. The change in magnetic diffraction  pattern between short to the long $0$--$\pi$ JJ limit \cite{Kirtley:IcH-PiLJJ} or the formation of spontaneous flux in the ground state of multiple $0$--$\pi$ phase boundaries in a long JJ could be studied \cite{ZenchukGoldobin04GS}. $0$--$\pi$ JJs may be used as the active part in the qubit, as was recently proposed \cite{Goldobin:2005:QuTu2Semifluxons}.\\

\section*{Acknowledgement}
The authors thank B. Hermanns for help with fabrication and E. Goldobin, R. Kleiner, D. Koelle and A. Ustinov for fruitful discussions. This work was supported by ESF program PiShift and Heraeus Foundation.


\begin{thebibliography}{10}

\bibitem{BuckelKleiner2004Superconductivity}
W.~Buckel and R.~Kleiner, {\em Superconductivity. Fundamentals and
  Applications.}, Wiley-VCH (2004).

\bibitem{Gurvitch82NbAlONb}
M.~Gurvitch, M.~A. Washington, H.~A. Huggins and J.~M. Rowell, IEEE Trans.
  Magn.\textbf{ 19}, 791 (1983).

\bibitem{buzdin05RMP}
A.~I. Buzdin, Rev. Mod. Phys.\textbf{ 77}, 935 (2005).

\bibitem{RoccaAprili05classicalspins}
M.~L.~Della Rocca, M.~Aprili, T.~Kontos, A.~Gomez and P.~Spatkis, Phys. Rev.
  Lett.\textbf{ 94}, 197003 (2005).

\bibitem{FrolovRyazanovSemifluxon}
S.~M. Frolov, D.~J.~Van Harlingen, V.~V. Bolginov, V.~A. Oboznov and V.~V.
  Ryazanov, Phys. Rev. B\textbf{ 74}, 020503 (2006).

\bibitem{WeidesFractVortex}
M.~Weides, M.~Kemmler, H.~Kohlstedt, R.~Waser, D.~Koelle, R.~Kleiner and E.~Goldobin, Phys. Rev.
  Lett.\textbf{ 97}, 247001 (2006).

\bibitem{Josephson}
B.~D. Josephson, Phys. Lett.\textbf{ 1}, 251 (1962).

\bibitem{Bulaevskii1977}
L.~Bulaevskii, V.~Kuzii and A.~Sobyanin, JETP Lett.\textbf{ 25}, 7 (1977).

\bibitem{Buzdin1991}
A.~I. Buzdin and M.~Yu. Kupriyanov, JETP Lett.\textbf{ 53}, 321 (1991).

\bibitem{Ryazanov2000}
A.~V. Veretennikov, V.~V. Ryazanov, V.~A. Oboznov, A~Yu. Rusanov, V.~A. Larkin
  and J.~Aarts, Physica B\textbf{ 284}, 495 (2000).

\bibitem{Blum:2002:IcOscillations}
Y.~Blum, A.~Tsukernik, M.~Karpovski and A.~Palevski, Phys. Rev. Lett.\textbf{
  89}, 187004 (2002).

\bibitem{Kontos02Negativecoupling}
T.~Kontos, M.~Aprili, J.~Lesueur and X.~Grison, Phys. Rev. Lett.\textbf{ 89},
  137007 (2002).

\bibitem{WeidesHighQualityJJ}
M.~Weides, M.~Kemmler, E.~Goldobin, D.~Koelle, R.~Kleiner, H.~Kohlstedt and
  A.~Buzdin, Appl. Phys. Lett.\textbf{ 89}, 122511 (2006).

\bibitem{Russo1978NonuniformJc}
M.~Russo and R.~Vaglio, Phys. Rev. B\textbf{ 17}, 2171 (1978).

\bibitem{Schwidal1969}
K.~Schwidal and R.~D. Finnegan, J. Appl. Phys.\textbf{ 40}, 213 (1969).

\bibitem{Barone1977PhysStat}
A.~Barone, G.~Patern\`{o}, M.~Russo,  and R.~Vaglio, Phys. Stat. Solidi
  (a)\textbf{ 41}, 393 (1977).

\bibitem{McLaughlinScottPRB1978}
D.~W. McLaughlin and A.~C. Scott, Phys. Rev. B\textbf{ 18}, 1652 (1978).

\bibitem{Vystavkin1988}
A.~N. Vystavkin, Yu.~F. Drachevskii, V.~P. Koshelets and I.~L. Serpuchenko,
  Sov. J. Low Temp. Phys.\textbf{ 14}, 357 (1988).

\bibitem{MaloUstinovJAP1990}
B.~A. Malomed and A.~V. Ustinov, J. Appl. Phys.\textbf{ 67}, 3791 (1990).

\bibitem{FistulPRB1999}
M.~V. Fistul, P.~Caputo and A.~V. Ustinov, Phys. Rev. B\textbf{ 60}, 13152
  (1999).

\bibitem{Lazarides2005}
N.~Lazarides, Supercond. Sci. Technol.\textbf{ 18}, 73 (2005).

\bibitem{LazaridesPeriodicDefects2003}
N.~Lazarides, Phys. Rev. B\textbf{ 68}, 92506 (2003).

\bibitem{GolubovUstinovPLA1988}
A.~Golubov, A.~Ustinov and I.~L. Serpuchenko, Phys. Lett. A\textbf{ 130}, 107
  (1988).

\bibitem{ItzlerTinkhamPRB95}
M.~A. Itzler and M.~Tinkham, Phys. Rev. B\textbf{ 51}, 435 (1995).

\bibitem{Bulaevskii1978}
L.~N. Bulaevskii, V.~V. Kuzii and A.~A. Sobyanin, Solid State Commun.\textbf{
  25}, 1053 (1978).

\bibitem{Xu:SF-shape}
J.~H. Xu, J.~H. Miller and C.~S. Ting, Phys. Rev. B\textbf{ 51}, 11958 (1995).

\bibitem{Goldobin02SF}
E.~Goldobin, D.~Koelle and R.~Kleiner, Phys. Rev. B\textbf{ 66}, 100508 (2002).

\bibitem{Tsuei:Review}
C.~C. Tsuei and J.~R. Kirtley, Rev. Mod. Phys.\textbf{ 72}, 969 (2000).

\bibitem{Kirtley:IcH-PiLJJ}
J.~R. Kirtley, K.~A. Moler and D.~J. Scalapino, Phys. Rev. B\textbf{ 56}, 886
  (1997).

\bibitem{Hilgenkamp:zigzag:SF}
H.~Hilgenkamp, Ariando, H.~J.~H. Smilde, D.~H.~A. Blank, G.~Rijnders,
  H.~Rogalla, J.~R. Kirtley and C.~C. Tsuei, Nature (London)\textbf{ 422}, 50
  (2003).

\bibitem{Goldobin03GS_biasReArrange}
E.~Goldobin, D.~Koelle and R.~Kleiner, Phys. Rev. B\textbf{ 67}, 224515 (2003).

\bibitem{ZenchukGoldobin04GS}
A.~Zenchuk and E.~Goldobin, Phys. Rev. B\textbf{ 69}, 024515 (2004).

\bibitem{Goldobin:2005:QuTu2Semifluxons}
E.~Goldobin, K.~Vogel, O.~Crasser, R.~Walser, W.~P. Schleich, D.~Koelle and
  R.~Kleiner, Phys. Rev. B\textbf{ 72}, 054527 (2005).

\bibitem{Goldobin04PRLDynamicsSF}
E.~Goldobin, A.~Sterck, T.~Gaber, D.~Koelle and R.~Kleiner, Phys. Rev.
  Lett.\textbf{ 92}, 057005 (2004).

\bibitem{WeidesFabricationJJPhysicaC}
M.~Weides, K.~Tillmann and H.~Kohlstedt, Physica C\textbf{ 437-438}, 349
  (2006).

\bibitem{SF6EtchingLichtenberger1993}
A.~W. Lichtenberger, D.~M. Lea and F.~L. Lloyd, IEEE Trans. Appl.
  Supercond.\textbf{ 3}, 2191 (1993).

\bibitem{WeidesPhD}
M.~Weides, Ph. D. thesis, Universit\"at zu K\"oln, Germany (2006).

\bibitem{BaronePaterno}
A.~Barone and G.~Paterno, {\em Physics and Applications of the Josephson
  Effect}, John Wiley \& Sons (1982).

\bibitem{Note0Pi}
The $I_c(h)$ characteristic for a $0$-$\pi$ JJ can be obtained in compact form from a general
expression valid for N $0$--$\pi$ phase boundaries, for $N=1$ and equal lengths of the $0$-
and $\pi$-parts, found in N.~Lazarides, Supercond. Sci. Technol.\textbf{ 17}, 585 (2004).

\bibitem{KirtleyImaging96PRL}
J.~R. Kirtley, C.~C. Tsuei, M.~Rupp, J.~Z. Sun, L.~S. Yu-Jahnes, A.~Gupta,
  M.~B. Ketchen, K.~A. Moler and M.~Bhushan, Phys. Rev. Lett.\textbf{ 76}, 1336
  (1996).

\end{thebibliography}
\end{document}